# pycnet-audio: A Python package to support bioacoustics data processing

Zachary J. Ruff[1,2]*, Damon B. Lesmeister[1,2]

[1] Department of Fisheries, Wildlife, and Conservation Sciences, Oregon State University, Corvallis, USA

[2] USDA Forest Service Pacific Northwest Research Station, Corvallis, USA

*Corresponding author. Email: ruffz@oregonstate.edu

## Summary

Passive acoustic monitoring is an emerging approach in wildlife research that leverages recent improvements in purpose-made automated recording units (ARUs). The general approach is to deploy ARUs in the field to record on a programmed schedule for extended periods (weeks or months), after which the audio data are retrieved. These data must then be processed, typically either by measuring or analyzing characteristics of the audio itself (e.g. calculating acoustic indices), or by searching for some signal of interest within the recordings,
e.g. vocalizations or other sounds produced by some target species, anthropogenic or environmental noise, etc. In the latter case, some method is required to locate the signal(s) of interest within the audio. While very small datasets can simply be searched manually, even modest projects can produce audio datasets on the order of $10^5$ hours of recordings, making manual review impractical and necessitating some form of automated detection.

pycnet-audio (Ruff 2024) is intended to provide a practical processing workflow for acoustic data, built around the PNW-Cnet model, which was initially developed by the U.S. Forest Service to support population monitoring of northern spotted owls (*Strix occidentalis caurina*) and other forest owls (Lesmeister and Jenkins 2022; Ruff et al. 2020). PNW-Cnet has been expanded to detect vocalizations of ca. 80 forest wildlife species and numerous forms of anthropogenic and environmental noise (Ruff et al. 2021, 2023).

## Statement of need

pycnet-audio is targeted at researchers and wildlife biologists who want to process acoustic data to extract meaningful ecological information in a logical and time-efficient way. Typically, researchers are interested in detecting sounds produced by one or more species of interest; these detections are then treated as indicators of species presence, as an index of vocal activity, etc. These parameters in turn can be modeled as a function of environmental variables to produce insight into the ecology of the species in question.

The challenge of automating the detection of vocalizing species in audio data has been studied extensively, and several software solutions exist. A major utility of the pycnet-audio package is that it is designed around the PNW-Cnet model and represents a standardization of processing tools and techniques used in northern spotted owl population monitoring and other biological monitoring efforts. The PNW-Cnet model has previously been made available through the Shiny_PNW-Cnet desktop app, which runs through RStudio and provides a graphical user interface (GUI) and various utilities for audio processing (Ruff et al. 2021). pycnet-audio is intended as a more flexible solution; it does not provide a GUI but does provide an application programming interface (API) specification, which allows users to define their own workflows and to easily script the processing of large datasets. Additionally, compared to Shiny_PNW-Cnet, pycnet-audio has fewer dependencies and a more streamlined installation procedure which leverages standard package management tools, e.g. pip.

PNW-Cnet is a fairly broad model trained on classes representing a range of taxa, with training examples ($n$=700,506 images) drawn primarily from real field recordings collected in forests of the Pacific Northwest. The design of pycnet-audio acknowledges its place within a larger bioacoustics workflow that encompasses initial data wrangling and organization, classification of the audio using an automated detection model, filter- ing raw model output to obtain a set of apparent detections for one or more species or sonotypes of interest, manual review and confirmation of these apparent detections, and subsequent analyses. pycnet-audio is designed to make it feasible to execute the data processing portion of this workflow on datasets of realistic size ($10^5$ hours or more) on consumer-grade hardware.

pycnet-audio is new software and has not yet been used in published research; however, PNW-Cnet has been used in several published studies, e.g. Appel et al. (2023), Rugg, Jenkins, and Lesmeister (2023), Duarte et al. (2024), and Weldy et al. (2023), demonstrating the value of the tools provided by this package in generating meaningful ecological insights.

# Overview

pycnet-audio is structured as a Python package containing a main importable module called pycnet and several submodules, which collectively define Python functions and classes for constructing processing work- flows for acoustic data. The package also provides command-line tools which can be utilized directly to process acoustic data. Generally, processing tasks are executed using the pycnet command from a command prompt program. This command takes two required arguments: mode, an indication of the processing task(s) to be performed; and target directory, a path to a directory containing audio data to be processed. Ad- ditionally, the pycnet command accepts a number of optional arguments which can be used to configure processing behavior.

Installation of pycnet-audio follows the standard procedure for Python packages and can be as simple as running pip install pycnet-audio from a command prompt. The only dependency which is not it- self a Python package is SoX, which pycnet-audio uses primarily for generating spectrograms and other miscellaneous audio processing tasks.

pycnet-audio is under active development, with new features and capabilities being added on an ongoing basis. We welcome contributions and feedback from fellow researchers and other interested parties.

# Acknowledgements

Funding was provided by USDA Forest Service and USDI Bureau of Land Management. The authors wish to thank those who have been involved in the development of the PNW-Cnet model, particularly the technicians who contributed many thousands of annotations to the training and testing datasets. The findings and conclusions in this publication are those of the authors and should not be construed to represent any official
U.S. Department of Agriculture determination or policy. The use of trade or firm names in this publication is for reader information and does not imply endorsement by the U.S. Government of any product or service.

# References

Appel, Cara L., Damon B. Lesmeister, Adam Duarte, Raymond J. Davis, Matthew J. Weldy, and Taal Levi. 2023. "Using Passive Acoustic Monitoring to Estimate Northern Spotted Owl Landscape Use and Pair Occupancy." *Ecosphere* 14 (2). https://doi.org/10.1002/ecs2.4421.

Duarte, Adam, Matthew J. Weldy, Damon B. Lesmeister, Zachary J. Ruff, Julianna M. A. Jenkins, Jonathon
J. Valente, and Matthew G. Betts. 2024. "Passive Acoustic Monitoring and Convolutional Neural Networks Facilitate High-Resolution and Broadscale Monitoring of a Threatened Species." *Ecological Indicators* 162 (May): 112016. https://doi.org/10.1016/j.ecolind.2024.112016.

Lesmeister, Damon B., and Julianna M. A. Jenkins. 2022. "Integrating New Technologies to Broaden the Scope of Northern Spotted Owl Monitoring and Linkage with USDA Forest Inventory Data." *Frontiers in Forests and Global Change* 5 (October): 966978. https://doi.org/10.3389/ffgc.2022.966978.

Ruff, Zachary J. 2024. "Pycnet-Audio: A Python Package for Bioacoustics Data Processing." *GitHub Repos- itory*. GitHub. https://github.com/zjruff/pycnet-audio.

Ruff, Zachary J., Damon B. Lesmeister, Cara L. Appel, and Christopher M. Sullivan. 2021. "Workflow and Convolutional Neural Network for Automated Identification of Animal Sounds." *Ecological Indicators* 124 (May): 107419. https://doi.org/10.1016/j.ecolind.2021.107419.

Ruff, Zachary J., Damon B. Lesmeister, Leila S. Duchac, Bharath K. Padmaraju, and Christopher M. Sulli- van. 2020. "Automated Identification of Avian Vocalizations with Deep Convolutional Neural Networks." *Remote Sensing in Ecology and Conservation* 6 (1): 79–92. https://doi.org/10.1002/rse2.125.

Ruff, Zachary J., Damon B. Lesmeister, Julianna M. A. Jenkins, and Christopher M. Sullivan. 2023. "PNW-Cnet V4: Automated Species Identification for Passive Acoustic Monitoring." *SoftwareX* 23 (July): 101473. https://doi.org/10.1016/j.softx.2023.101473.

Rugg, Natalie M., Julianna M. A. Jenkins, and Damon B. Lesmeister. 2023. "Western Screech-Owl Occu- pancy in the Face of an Invasive Predator." *Global Ecology and Conservation* 48 (December): e02753. https://doi.org/10.1016/j.gecco.2023.e02753.

Weldy, Matthew J., Damon B. Lesmeister, Charles B. Yackulic, Cara L. Appel, Chris McCafferty, and J. David Wiens. 2023. "Long-Term Monitoring in Transition: Resolving Spatial Mismatch and Integrating Multistate Occupancy Data." *Ecological Indicators* 146 (February): 109815. https://doi.org/10.1016/j. ecolind.2022.109815.